\documentclass[
authoryear
,11pt
,a4paper
,helvet
,final
]{elsarticle}

\usepackage[english]{babel}
\usepackage[T1]{fontenc}
\usepackage{amssymb,amstext,amsmath,amsfonts}
\usepackage{graphicx}
\usepackage{multirow}
\usepackage{url}
\usepackage{color}
\usepackage{longtable}
\usepackage[hmargin=2cm,vmargin=1.5cm]{geometry}
\usepackage{fancyhdr}

\newcommand{\be}{\begin{equation}}
\newcommand{\ee}{\end{equation}}
\newcommand{\bd}{\begin{displaymath}}
\newcommand{\ed}{\end{displaymath}}
\newcommand{\BE}{\begin{eqnarray}}
\newcommand{\EE}{\end{eqnarray}}

\usepackage[
 breaklinks=false
 ,colorlinks=true
 ,linkcolor=blue
 ,citecolor=blue
 ,urlcolor=black
 ,backref=none,
 ,pagebackref=false
 ,hyperfootnotes=false
 ]{hyperref} 
\usepackage{xspace}
\usepackage{helvet}
\usepackage{setspace}
\usepackage{typearea} 
\usepackage{placeins}
\usepackage[active]{srcltx}

\date{July 2012}

\newcommand{\e}{\text{e}}

\newcommand{\eq}[1]{(\ref{eq:#1})}

\begin{document}

\begin{frontmatter}
\sffamily

\title{The mechanics of stochastic slowdown in evolutionary games}
\author[1]{Philipp M.~Altrock\corref{cor1}}
\ead{altrock@evolbio.mpg.de}
\cortext[cor1]{Corresponding author}
\author[1]{Arne Traulsen}
\ead{traulsen@evolbio.mpg.de}
\address[1]{\small Research Group for Evolutionary Theory, Max-Planck-Institute for Evolutionary Biology, August-Thienemann-Str. 2, D--24306 Pl\"on, Germany
}
\author[2]{Tobias Galla}
\ead{Tobias.Galla@manchester.ac.uk}
\address[2]{Theoretical Physics, School of Physics \& Astronomy, The University of Manchester, Manchester M 13 9PL, UK}

\begin{abstract}
\sffamily
We study the stochastic dynamics of evolutionary games, and focus on the so-called `stochastic slowdown' effect, previously observed in \citep{altrock:2010aa} for simple evolutionary dynamics. Slowdown here refers to the fact that a beneficial mutation may take longer to fixate than a neutral one. More precisely, the fixation time conditioned on the mutant taking over can show a maximum at intermediate selection strength. We show that this phenomenon is present in the Prisoner's Dilemma, and also discuss counterintuitive slowdown and speedup in coexistence games. In order to establish the microscopic origins of these phenomena, we calculate the average sojourn times. This allows us to identify the transient states which contribute most to the slowdown effect, and enables us to provide an understanding of slowdown in the takeover of a small group of cooperators by defectors in the Prisoner's Dilemma: Defection spreads fast initially, but the final steps to takeover can be delayed substantially. The analysis of coexistence games reveals even more intricate non-monotonic behavior.  In small populations, the conditional average fixation time can show multiple extrema as a function of the selection strength, e.g., slowdown, speedup, and slowdown again. We classify generic $2\times2$ games with respect to the possibility to observe non-monotonic behavior of the conditional average fixation time as a function of selection strength. 
\end{abstract}

\end{frontmatter}

\newpage

\pagestyle{fancyplain}
\fancyhf{}
\chead{\fancyplain{}{\sffamily The mechanics of stochastic slowdown in evolutionary games}}
\rfoot{\fancyplain{}{\sffamily \thepage}}
\lfoot{\fancyplain{}{\sffamily P.M.~Altrock, A.~Traulsen, and T.~Galla}}

\sffamily

\section{Introduction}\label{sec:intro}

The theory of evolutionary games describes the effects of selection, reproduction and mutation in competitive environments of interacting agents. 
In an evolutionary setting, reproductive success (or fitness) depends on the performance in the evolutionary game. 
The strategies, or types of the game are assumed to be hard-wired to an individual's genotype, and passed on from parent to offspring. 
Natural selection acts, such that more successful strategies spread faster over time than less successful strategies. 
Whether a particular strategy is successful or not depends on the state of the overall population \citep{maynard-smith:1973to,taylor:1978wv,hofbauer:1979mm,zeeman:1980ze,nowak:2004aa}. 
An individual's fitness will generally depend on the frequencies of all strategies present in the population. 
This is referred to as `frequency dependent selection' \citep{maynard-smith:1982to,weibull:1995hp,hofbauer:1998mm,nowak:2006bo}. 
For large populations, such systems are traditionally described by deterministic equations of motion, typically replicator dynamics or close variations \citep{schuster:1983le,hofbauer:1998mm,gintis:2000bv,sandholm:2010bo}. We here focus instead on the stochastic evolutionary dynamics of two strategies, a mutant type $A$ and a wildtype $B$, in a finite population of size $N$.

A significant body of literature is now available on the stochastic dynamics of evolutionary processes, and a number of phenomena induced by stochasticity have now been identified \citep{nowak:2004pw,antal:2006aa,altrock:2010aa,traulsen:2006bb,traulsen:2005hp,traulsen:2009bb,claussen:2005eh,claussen:2007aa,cremer:2008aa,bladon:2010ws}.  Once all individuals of a given type have been eliminated, they are never re-introduced in the absence of mutation. Naturally, a stochastic evolutionary process without mutations ends up in such an absorbing state. 
If selection is sufficiently weak, this allows the fixation of disfavored types, i.e., the possibility that a stochastic evolutionary process ends up in configurations consisting of a type which does not have the highest fitness \citep{nowak:2004pw}. 
Fixation will ultimately also occur in coexistence games, where the replicator dynamics predicts a mixed asymptotic state. However, in this case the time-to-fixation diverges with increasing selection \citep{traulsen:2006ab,traulsen:2007aa}.

Here, we analyze the phenomenon of `stochastic slowdown' observed in games between two types, a wildtype and a mutant \citep{altrock:2010aa}. 
Assuming an evolutionary game in which the mutant is always better off than the wildtype (no matter what the configuration of the population is), it was shown that increasing the selection pressure (and with it the evolutionary advantage of the mutant) fixation into the all-mutant state can slow down.
In other words, fixation can take longer on average at non-zero selection strength than in the neutral case of no selection at all. 
Consequently, a beneficial mutation can take longer to take over a population than a neutral one. 
This somewhat counter-intuitive effect was investigated in \citep{altrock:2010aa}, where a number of relatively simple evolutionary setups are analyzed and constraints on parameters such as the population size required to observe this kind of stochastic slowdown are derived.  

Our goal here is to extend these observations to the case of evolutionary games. Specifically we focus on the Prisoner's Dilemma game and on coexistence games. 
We provide an in-depth analysis of the transient dynamics of stochastic slowdown, and investigate what states of the system contribute most to the slowdown effect. 
To this end, we calculate the average conditional sojourn times, i.e.~the average time the system spends in any particular configuration before fixating into the all-mutant state, and provide a weak selection approximation.

In Section \ref{sec:model} we describe the general setup of evolutionary dynamics in finite populations with two types, and define the quantities of interest, in particular fixation and sojourn times. In Section \ref{sec:Toy}, we introduce a minimal model in which slowdown can be observed and analyzed, which allows us to develop an intuitive understanding of stochastic slowdown. In Section \ref{sec:GameDyn} we introduce the model describing interactions between individuals of two different types in the setting of an evolutionary game. Section \ref{sec:res} contains our main results. Further discussion and conclusions can be found in Section \ref{sec:conc}. Some of the technical details of our analysis can be found in the Appendix. Table \ref{tab:MainSymbols} gives a list of symbols used in the main part of this article, Table \ref{tab:ToySymbols} gives a list of symbols exclusively used in Section \ref{sec:Toy}. 

{\footnotesize
\begin{longtable}{p{0.125\textwidth}p{0.8\textwidth}}
	\hline\hline 	
	Symbol & Definition\\\hline 
	$A$, $B$ & Types (strategies) in the population. $A$ is the invading mutant, $B$ is the wildtype. \\
	$N$	& Population size.\\
	$i$, $j$ & Number of individuals of type $A$, also called state of the system. States $0$ and $N$ are absorbing.\\ 
	$T_i^+$ & Transition probability from state $i$ to state $i+1$.\\
	$T_i^-$ & Transition probability from state $i$ to state $i-1$.\\
	$P_t(j)$ & Probability that the population is in state $j$ at time $t$. Time is discrete.\\
	$P_t(j|i)$ & Conditional probability that the population is in state $j$ at time $t$ when it was in state $i$ at time 0.\\
	$p_t(j|i)$ & Conditional probability to {\em \sffamily enter} state $j$ at time $t$, when it started from state $i$ at time 0.\\
	$\phi_{i j}$ &	Probability to ever visit state $j$, starting in state $i$.\\
	$\phi_{i}$ &	Fixation probability in state $N$, starting in state $i$, $\phi_i=\phi_{i N}$, and $1-\phi_i=\phi_{i 0}$.\\
	$r_i$ &	Probability to ever return to state $i$, once there.\\
	$q_t(j|i)$	&	Conditional probability of time $t$ spent in state $j$, starting from $i$.\\
	$t_{i j}$	& Average sojourn time in $j$, starting in $i$: First moment of $q_t(j|i)$, $t_{i j}=\sum_{t=1}^\infty t\,q_t(j|i)$.\\
	$t_i$	&	Unconditional average fixation time starting from $i$.\\
	$t_i^0$, $t_i^N$	&	Conditional average fixation time into $0$ or $N$, respectively, starting from $i$.\\
	$a$, $b$, $c$, $d$	&	Payoffs of a single interaction between individuals.\\
	$\pi_A$, $\pi_B$	&	Average payoffs of $A$, $B$ in a well-mixed population of fixed size $N$.\\
	$u$, $v$	&	Parameters that characterize the evolutionary game in state $i$. $\pi_A-\pi_B=u\,i +v$.\\
	$\beta$	&	Selection intensity. $\beta=0$ means neutral evolution.\\
	$f_A$, $f_B$	&	Fitness of $A$, $B$, e.g., $f_A=\e^{\beta\,\pi_A}$.\\
	$\bar f$	&	Expected fitness in the population. In state $i$ we have $\bar f=(i\,f_A+(N-i)f_B)/N$.\\
	$i^*$	&	Points where $T_{i^*}^+=T_{i^*}^-$, i.e.~$i^*=0,N,\,-u/v$.\\
	$g_i^+$, $g_i^-$	& Game-dependent part of the transition probabilities $T_i^{\pm}$ in state $i$.\\
	$\hat D_1$, $\hat D_2$, $\check D_1$	&	Coefficients of the weak selection expansion of $t_{i j}$ that only depend on $i$, $j$, $N$.\\
	$\tilde E_1$, $\tilde E_2$, $\hat E_1$, $\hat E_2$, $\check E_1$, $\check E_2$	&	Coefficients of the weak selection expansions of $t_i$, $t_i^0$, $t_i^N$ that only depend on $i$, $N$.\\
	$H_k$	&	Harmonic number, $H_k=\sum_{l=1}^k1/l$.\\
	\hline\hline 
	\caption{\sffamily Symbols used in this article, roughly in order of their appearance. Section \ref{sec:Toy} has its own table of symbols.}
	\label{tab:MainSymbols}
\end{longtable}
}
\section{Model}\label{sec:model}

\subsection*{\sffamily General Setup}

We model a well-mixed population of size $N$, subject to a birth-death process in discrete time of overlapping generations that keeps the size of the population constant. We consider the interaction of two types, a mutant $A$ and a wildtype $B$. We denote the number of mutants in the population by $i$, and accordingly the number of individuals of the wildtype is $N-i$.

The birth-death process is fully characterized by the probabilities to increase or decrease the number of individuals of the mutant type at each time step, i.e.~whether to increase $i$ to $i+1$ (a mutant displaces an individual of the wildtype) or whether to reduce it to $i-1$ (an individual of the wildtype displaces a mutant), or not to make any change at all. 
We denote the probability of a the occurrence of a transition from $i$ to $i+1$ in a given time step by $T_i^+$, and $T_i^-$ accordingly denotes the probability that a transition from $i$ to $i-1$ occurs. The quantity $1-T_i^+-T_i^-$ is then the probability that the population remains in state $i$. We exclude mutations such that the states $i=0$ and $i=N$ are absorbing, i.e., we have $T_0^+=T_N^-=0$. The dynamics of the system is then governed by the following master equation
\be
P_{t+1}(j)=(1-T_j^+-T_j^-)P_t(j)+T_{j-1}^+P_t(j-1)+T_{j+1}^-P_t(j+1),
\ee
where $P_t(j)$ denotes the probability of finding the system in state $j$ at time $t$. Similarly paths can be conditioned on an initial state $i$. Writing $P_t(j|i)$ for the conditional probability of finding the system in state $j$ exactly $t$ time steps after starting the dynamics in state $i$, we have
\be
P_{t+1}(j|i)=(1-T_j^+-T_j^-)P_t(j|i)+T_{j-1}^+P_t(j-1|i)+T_{j+1}^-P_t(j+1|i).
\ee

\subsection*{\sffamily Fixation Probabilities}
If the system is started from a configuration with $i$ individuals of the mutant type, 
then it will eventually fixate in one of the two absorbing states at $i=0$ or $i=N$. The probability that this fixation occurs at $i=N$ (the mutant type takes over the entire population) is denoted by $\phi_i$, where $i$ is the initial number of mutants. 
Accordingly, fixation at $i=0$ (mutant type goes extinct) occurs with probability $1-\phi_i$. 

In order to calculate fixation probabilities it is useful to first consider the probabilities $p_t(j|i)$ describing the event that the system {\em \sffamily enters} state $j$ exactly $t$ time steps after having been started in state $i$. We stress that this is not the same as the probability $P_t(j|i)$ of {\em \sffamily being found} in state $j$ after a lag period of $t$ time steps and having started at $i$. Instead, $p_t(j|i)$ only captures events in which the system was in a state different from $j$ at time step $t-1$ after starting from $i$. One then has the following backward master equation \citep{goel:1974aa,karlin:1975xg}
\be\label{eq:Master01b}
p_{t+1}(j|i)=(1-T_i^+-T_i^-)p_t(j|i)+T_i^+p_t(j|i+1)+T_i^-p_t(j|i-1),
\ee
where again $i$ denotes the initial state, and $j$ the state $t$ time steps later. Summing over all $t$ yields the recursion 
\be\label{eq:Master02}
\phi_{i j}=(1-T_i^+-T_i^-)\phi_{i j}+T_i^+\phi_{i+1 j}+T_i^-\phi_{i-1 j}.
\ee
Here $\phi_{ij}=\sum_{t=0}^\infty p_t(j|i)$ is the probability that the system reaches state $j$ at any later time if the system is started in state $i$.
The solution of this recursion is given in Appendix \ref{sec:App01}. 
The fixation probability of a group of $i$ mutants of type $A$ is $\phi_i=\phi_{ i N}$.
Similarly, $1-\phi_i=\phi_{ i 0}$ is the probability that the mutation goes extinct in the population if there are $i$ mutants initially. Once the $p_t(j|i)$ are known, it is straightforward to compute the average time to fixation, conditioned for example on cases in which the mutant takes over. The average conditional fixation time is given by 
$\sum_{t=0}^\infty t \,p_t(N|i)/\phi_i$ \citep{antal:2006aa,traulsen:2009bb}. 
Recall here that $p_t(N|i)$ denotes the probability that the system reaches state $N$ precisely $t$ time steps after starting in state $i$. More generally, the $k^{\text{th}}$-moment of the conditional fixation times can be obtained as $\sum_{t=0}^\infty t^k\,p_t(N|i)/\phi_i$. 

\subsection*{\sffamily Sojourn Times}

Sojourn times are a helpful tool with which to characterize the transient dynamics of stochastic processes with fixation \citep{ohtsuki:2007aa}. The sojourn time $t_{i j}$ is the average total time a population started in state $i$ spends in the state with $j$ mutants before absorption. This includes returns to state $j$ and time steps in which the system remains in $j$, i.e., $t_{ij}$ is the average {\em \sffamily total} number of time steps spent in $j$ until absorption if started from $i$. 
 
 {\em \sffamily Unconditional sojourn times:} 
We can identify the sojourn times by considering the escape process from each of the internal states $i=1,\dots,N-1$ \citep{ewens:2004qe}.  
Once in $i$ at time step $t$,  the probability $r_i$ that the process is found in state $i$ again at any future time step $t'>t$ is given by
\begin{align}\label{eq:ReturnProb}
\begin{split}
	r_i=(1-T_i^+-T_i^-)+T_i^+\phi_{i+1 i}+T_i^-\phi_{i-1 i}.
\end{split}
\end{align} 
Then, the conditional probability for the system starting in state $i$ to spend a total time of $t>0$ steps in state $j$ before absorption is 
\begin{align}\label{eq:Sojourn02}
	q_t(j|i)=\phi_{i j}\,r_j^{t-1}(1-r_j).
\end{align}
The sojourn time at $j$, conditioned on a starting point $i$ is obtained as the first moment of this distribution
\begin{align}\label{eq:Sojourn03}
	t_{i j} = \phi_{i j}(1-r_j)\,\sum\limits_{t=1}^{\infty}t\,r_j^{t-1}
		= \frac{\phi_{i j}}{1-r_j}. 
\end{align}
 This expression can be simplified further, noting that $r_i=1+T_i^+(\phi_{i+1 i}-1)+T_i^-(\phi_{i-1 i}-1)$. 
Consequently, the average sojourn time in state $j$, starting in $i$, reads
\begin{align}\label{eq:Uncond01}
	t_{i j} = \frac{\phi_{i j}} {T_j^+(1-\phi_{j+1 j})+T_j^-(1-\phi_{j-1 j})}. 
\end{align}
 The average life-time $t_i$, i.e., the unconditional fixation time (see, e.g., \cite{altrock:2009nj}) with initial condition $i$ is then given by the sum over all average sojourn times, $t_i=\sum_{j=1}^{N-1}t_{i j}$.  

{\em \sffamily Conditional sojourn times:} We will now consider average sojourn and fixation times, conditioned on paths ending in the all-mutant state.
We can here use general results relating conditional to unconditional quantities, see \citep{ewens:2004qe}. If we denote sojourn times conditioned on fixation on the all-mutant state by $t_{ij}^N$ then one has $t_{ij}^N=\phi_j/\phi_it_{ij}$. 
The conditional fixation time $t_i^{N}$ follows as
\begin{align}\label{eq:Cond01}
	t_i^N=\sum\limits_{j=1}^{N-1}\frac{\phi_j}{\phi_i}\,t_{i j}=\sum\limits_{j=1}^{N-1}\frac{\phi_j}{\phi_i}\frac{\phi_{i j}}{T_j^+(1-\phi_{j+1 j})+T_j^-(1-\phi_{j-1 j})}.
\end{align}

We note that the average time to fixation conditioned on arrival in the state $0$ (i.e.~a population free of mutants) can be obtained in an analogous manner as
\be
t_i^0=\sum_{j=1}^{N-1}\frac{1-\phi_j}{1-\phi_i}\,t_{i j}.
\ee
The expressions for average fixation and sojourn times are used in Section \ref{sec:res} to understand the statistical mechanics of trajectories conditioned on fixation of the mutant based on mutant-wildtype interactions cast into an evolutionary game. Before we get to these results we discuss a simpler model capturing the essence of probabilistic fixation of an advantageous mutant. The benefit of this excursion is a better understanding of the slowdown effect that we discuss for the evolutionary Prisoner' s Dilemma game.

\section{Simplified Model for Stochastic Slowdown}\label{sec:Toy}

\begin{figure}
\begin{center}
\includegraphics[width=.9\linewidth]{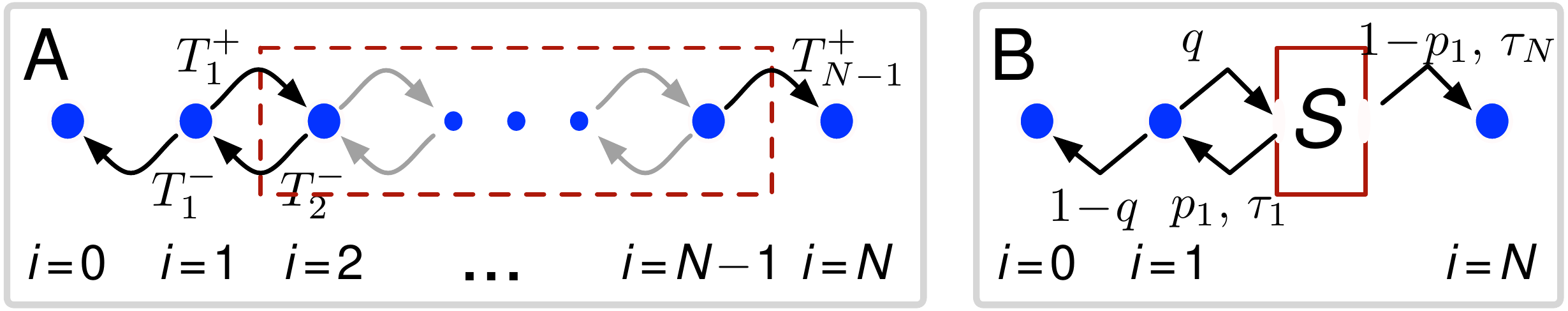}
\end{center}
\caption{\sffamily Illustration of the birth-death model of a population of size $N$ (panel {\bf \sffamily A}), characterized by the transition probabilities in each state, $T_i^\pm$. Panel {\bf \sffamily B} shows the simplified model in which states $i=2,\dots,N-1$ are aggregated into a coarse-grained state $S$ characterized by the variables $p_1,\tau_1$ and $\tau_N$ (see text for details).
}\label{fig:boxmodel}
\end{figure}

In order to develop a first intuitive understanding of the stochastic spread of a beneficial mutation let us consider a simplified model of the birth-death process. In order to disentangle the different effects contributing to the occurrence of slowdown, we focus on a basic model with only a small number of possible states. The left-hand panel of Figure \ref{fig:boxmodel} shows the setup of a general birth-death process in a population of $N$ individuals. We now take a coarse-grained view of the system and collect all states $i=2,3,\dots,N-1$ into a subsystem $S$, as shown in the right-hand panel of the figure. The birth-death process can be thought of the hopping dynamics between four different states, $i=0$, $i=1$, $S$ and $i=N$, where $i=0$ and $i=N$ are absorbing. 
State $i=1$ is characterized fully by $T_1^+$ and $T_1^-$. 
These also determine the probability $q=T_1^+/(T_1^++T_1^-)$ with which the next move of the system (out of state $1$) is into state $S$ and not into state 0. 
The average time the system stays in state $i=1$ is $s_1=1/(T_1^++T_1^-)$. 
Interestingly, this is also the average waiting time conditioned on the event that the next move is to state $S$ as well as the average waiting time conditioned on the event that the next move is to state $i=0$ \citep{antal:2006aa,taylor:2006jt}. 

We assume now that the system has entered state $S$ at time $t$. The previous state at $t-1$ then must have been $i=1$. In this situation state $S$ is characterized by the following three quantities: 
(i) The probability $p_1$  that the next state after leaving $S$ is $i=1$; 
(ii) the average time $\tau_1$ that elapses between entering state $S$ and returning to state $i=1$;
(iii) the average time $\tau_N$ that elapses between entering state $S$ and reaching state $i=N$. This is illustrated in the right-hand panel of Figure \ref{fig:boxmodel}. We stress that the quantities $T_i^\pm$ depicted in the left-hand panel of Figure \ref{fig:boxmodel} are the probabilities with which a birth-death event occurs in the {\em \sffamily next} time step. The quantity $1-T_i^+-T_i^-$ is the probability that no event occurs in the next step, which may well be positive. The quantity $p_1$ illustrated in the right-hand panel of the figure is the probability that the next hop at {\em \sffamily any} time is into state $i=1$ if the system is in $S$, the next hop will be towards state $N$ with probability $1-p_1$. The difference between the two cases is indicated by the two types of arrows in Figure \ref{fig:boxmodel}.

We now address the average conditional fixation time $t_1^N$. 
Assume that the system starts in state $i=1$. 
It can then reach $i=N$ by first hopping to state $S$ and then to $i=N$ without returning to $i=1$. 
This will occur with probability $\rho(0)=q(1-p_1)$ and will on average take $s_1+\tau_N$ time steps. 
Alternatively, the system may enter $S$, return to $i=1$ precisely once, then return to $S$ and then to $i=N$. 
The probability for this event is $\rho(1)=q(p_1 q)(1-p_1)$ and it will require $2s_1+\tau_1+\tau_N$ time steps. Analogously, the probability that exactly $k$ returns to state $i=1$ occur before absorption at $i=N$ occurs is
\be
 \rho(k)=q(p_1q)^k(1-p_1).
 \ee 
 The time required for this trajectory is $s_1+\tau_N+k(s_1+\tau_1)$. The probability for the particle to end up at $N$ eventually (and not at $i=0$) is then
 \be\label{eq:phi}
 \phi_1=\sum_{k=0}^\infty \rho(k)=\frac{q(1-p_1)}{1-p_1q}.
 \ee
 The conditional probability that the particle returns to site $i=1$ exactly $k$ times given that it is eventually absorbed at $N$ is then
 \be
 \rho^N(k)=\frac{\rho(k)}{\phi_1}=(p_1q)^k(1-p_1q).
 \ee
Averaging the time required for these processes over $k$, we find for the average conditional fixation time starting at $i=1$
\BE
t_1^N=\sum_k \rho^N(k)\left[s_1+\tau_N+k(s_1+\tau_1)\right]=s_1+\tau_N+(s_1+\tau_1)\frac{p_1q}{1-p_1q}.
\EE
With this, we can study how the properties $s_1$ and $q$ of state $i=1$ and those of the coarse-grained state $S$ (i.e., $\tau_1,\tau_N,p_1$) contribute to the overall fixation time. 
Consider for example the simple model system given by $N=4$ and
\BE
&T_1^+=\frac{1+\beta}{2}, ~ T_1^-=\frac{1-\beta}{2} \\
&T_2^+=T_2^-=T_3^+=T_3^-=\frac{1}{2}
\EE
where $0\leq\beta\leq 1$ introduces a net bias towards moving into $S$ when positioned at $i=1$. This is a special case of the constant-step process analyzed in \citep{altrock:2010aa}, one of the most basic models exhibiting the phenomenon of stochastic slowdown. 
In this model, we have $T_1^++T_1^-=1$ independently of $\beta$, so that $s_1$ does not depend on $\beta$. At the same time the subsystem $S$ is composed of sites $i=2$ and $i=3$, and the corresponding transition rates $T_2^\pm$ and $T_3^\pm$ are independent of $\beta$. The only relevant model variable affected by changes in $\beta$ is $q=(1+\beta)/2$ so that $t_1^N$ increases as $\beta$ is increased (note here that the term $p_1q/(1-p_1q)$ is increasing and highly non-linear in $q$). Thus, the fixation time $t_1^N$ increases with $\beta$, despite the fact that with increasing $\beta$ the system tends to move towards $N$. 

Intuitively, the slowdown effect in this toy model can be understood as follows: 
Consider an ensemble of (independent) realizations of the birth-death process, all started at $t=0$ in state $i=1$. These realizations may be thought of as particles hopping about independently on the set of available states $i=0$, $i=1$, $S$ and $i=N$. At a given time $t>0$ a number of such particles will have been absorbed at $i=0$ and $i=N$, a number of particles will be located at $i=1$, and a further set of particles will be in state $S$. Whenever a particular particle leaves state $S$ and enters state $i=1$ it is returned (after some waiting time) into the pool of occupants of $S$ with probability $q$. With probability $1-q$ it is absorbed at $i=0$ (again after some waiting time). Increasing the model parameter $q$ and thus returning an increased fraction of such particles into the subsystem $S$ will (on average) increase the typical `age' of particles in the pool at state $S$. Given that all particles being absorbed at $i=N$ are drawn from this pool of particles in state $S$, this will in turn lead to an increased age among the particles arriving at $i=N$, and hence an increased conditional fixation time $t_1^N$. 
While this intuitive interpretation of stochastic slowdown is valid for the toy model of the constant-step process, the interplay between the different factors contributing to the conditional fixation time is more intricate in other birth-death processes, such as the frequency dependent Moran process. 
In the remainder of the paper we will study these effects in the context of evolutionary games. 
Furthermore, other counter-intuitive effects are found, including multiple extrema of the conditional fixation time as a function of selection strength.

{\footnotesize
\begin{longtable}{p{0.125\textwidth}p{0.775\textwidth}}
	\hline\hline 	
	Symbol & Definition\\\hline 
	$S$	&	Subsystem consisting of all states $i=2,...,N\!-\!1$.\\
	$q$	&	Conditional probability of a transition from $1$ into $S$, $q=T_1^+/(T_1^++T_1^-)$.\\
	$s_1$	&	Average time spent in state 1 between arrival and subsequent departure, $s_1=1/(T_1^++T_1^-)$.\\
	$p_1$	&	Conditional probability that the system leaves $S$ into $i=1$.\\
	$\tau_1$, $\tau_N$		&	Average times between entering $S$ and leaving again into 1,$N$.\\
	$\rho(k)$	&	Conditional probability of fixation in $0$, or $N$, given that the initial state 1 is visited exactly $k$ times.\\
	$\rho^N(k)$	&	$\rho^N(k)=\rho(k)/\phi_1$.\\
	$\beta$	&	Net bias towards moving from $i=1$ into $S$.\\
	\hline\hline 
	\caption{\sffamily Additional symbols used exclusively in Section \ref{sec:Toy}.}
	\label{tab:ToySymbols}
\end{longtable}
}

\section{Game Dynamics}\label{sec:GameDyn}

We consider frequency dependent selection in evolutionary games between two types $A$ and $B$. 
The payoff to each of the two interacting individuals is $a$ if they are both of type $A$, whereas 
two interacting individuals of type $B$ receive $d$ each upon mutual interaction.  An $A$ interacting with a $B$ receives $b$, whereas $B$ obtains $c$ in this situation. 
This defines a symmetric $2\times2$ game with the payoff matrix
\begin{align}\label{eq:Pmatrix}
\bordermatrix{
& A & B \cr
A & a & b \cr
B & c & d \cr}
\end{align}
We assume that the interaction between individuals occur on a much faster time scale than the birth-death dynamics, so that the payoff of an individual is given by the expected payoff obtained from interaction with a randomly chosen individual.  
These average payoffs are then given by 
\begin{align}
	\pi_A =&\,\frac{a(i-1)+b{(N-i)}}{N-1}\label{eq:AvPayoffA},\\
	\pi_B =&\,\frac{c\,i+d{(N-i-1)}}{N-1}\label{eq:AvPayoffB},
\end{align}
for players of types $A$ and $B$, respectively. 
We have here excluded self-interaction. It is worth mentioning that the average payoffs in such games are (affine) linear in the frequency, $i$, of player of type $A$.

In order to define the evolutionary process we need to specify how the payoffs for each of the two strategies determine the success in reproduction. Reproductive success is determined by fitness, which in turn is a function of the average payoff received by a particular strategy. 
We denote fitness for the two types of players by $f_A$ and $f_B$, respectively. Several choices are possible for the mapping from payoff, see Eqs. (\ref{eq:AvPayoffA},\ref{eq:AvPayoffB}), to fitness. Nature does not tell us what fitness function to apply, but, by definition, fitness is a monotonically increasing function of the payoff \citep{wu:2010aa}.   
Following \citep{traulsen:2008aa}, we choose  
\begin{align}
	 f_A=\exp(\beta \pi_A), \label{eq:fitnessA}\\
	 f_B=\exp(\beta \pi_B). \label{eq:fitnessB}
\end{align}
While this a special choice, the phenomena discussed here are also present for other choices, e.g., a linear payoff-to-fitness mapping. 
It is important to remember that payoff and fitness are both frequency dependent.
The parameter $\beta$, referred to as the intensity of selection, determines how strongly the payoff from an individual's interactions influences fitness. For vanishing selection intensity ($\beta=0$), evolution is neutral for any game, i.e., no strategy has an advantage over the other in any state of the population. In our setup we have $f_A=f_B=1$ for $\beta=0$. 

In order to define the actual dynamics of the systems we make use of the frequency dependent Moran process \citep{moran:1962ef,nowak:2004pw}. In this process the transition probabilities of the birth-death process are given by
\begin{align}
	T_i^+=&\frac{i}{N}\frac{(N-i)}{N}\frac{f_A}{\overline f}\label{eq:Trans01a},\\
	T_i^-=&\frac{i}{N}\frac{(N-i)}{N}  \frac{f_B}{\overline f}\label{eq:Trans01b},\\
	T_i^0=&\,1-T_i^+-T_i^-\label{eq:Trans01c},
\end{align}
where 
$
\overline f=\frac{i}{N}f_A+\frac{N-i}{N}f_B$
is the expected fitness of a randomly chosen individual in the population. We point out that this is a one-step process, so at each step of the dynamics at most one replacement of an individual by another occurs. Accordingly, the transition rates between states $i$ and $j$ vanish for all $|i-j|>1$.
  
The transition probabilities in Eqs.~(\ref{eq:Trans01a},\ref{eq:Trans01b},\ref{eq:Trans01c}) depend only on the ratio of the fitness values, $f_A/f_B$. 
Given the exponential fitness mapping, this ratio in turn only depends on the average payoff difference $\pi_A-\pi_B$. 
The same dependence on the payoff difference is also obtained for many other mappings in the case of linear weak selection 
\citep{bladon:2010ws,wu:2010aa}.
In our case, the payoff difference is linear in $i$ and it can hence be written in the form $\pi_A-\pi_B=u\,i+v$. For games of the specific form given in Eq.  (\ref{eq:Pmatrix}) we have 
\begin{align}
	u=&\,\frac{a+d-(b+c)}{N-1}\label{eq:PackageRule1},\\
	v=&\,b-d+\frac{b-a}{N-1}\label{eq:PackageRule2}. 
\end{align}
These two parameters govern the frequency dependent evolutionary dynamics. The frequency dependent term proportional to $u$ measures the cumulated success of pure interactions ($A$-$A$, and $B$-$B$) versus the success of mixed interactions ($A$-$B$, and $B$-$A$). The constant contribution is proportional to $v$, which roughly gives a measure for the `invasion barrier' of an $A$ mutant, i.e., the performance of $A$-$B$ versus $B$-$B$, minus a small correction from $A$-$B$ versus $A$-$A$ interactions. 

In addition, the roots of the gradient of selection, $T_i^+-T_i^-$, i.e., the points $i^*$ for which there is no net advantage to either strategy, can also be written in terms of $u$, and $v$. 
Solving $T^+_{i^*}-T^-_{i^*}=0$ leads to $i^*=0,N$, and $i^*=-v/u$. 
We point out that this latter point need not be an integer, and that it may lie outside the space of allowed configurations, i.e., we may have $i^*<0$ or $i^*>N$ -- in these cases $A$ is always advantageous over $B$ or vice versa. 
For $0<i^*<N$, we either have a coexistence game or a coordination game.

\section{Results and Discussion}\label{sec:res}

 \subsection*{\sffamily Neutral Evolution}

Neutral evolution is the natural benchmark case of a strategy's evolutionary success, both in population genetics and in evolutionary games in finite populations \citep{ewens:2004qe,nowak:2006bo}. 
Neutral evolution emerges in the special case of $\pi_A-\pi_B=0$ for any $i$ and $\beta$, or generally for $\beta=0$. 
This leads to $T_i^+=T_i^-$, and thus to 
\begin{align}\label{eq:Neutral01}
\begin{split}
	\phi_{ij}=\begin{cases}
		\frac{N-i}{N-j}\,\, \text{for}\,\, i>j\\
		\frac{i}{j}\,\, \text{for}\,\, i\leq j, 
	\end{cases}
\end{split}
\end{align}
see Appendix \ref{sec:App01}.
The neutral transition probabilities read
\begin{align}\label{eq:Neutral02}
\begin{split}
	T_i^\pm=\frac{i(N-i)}{N^2}, 
\end{split}
\end{align}
such that the probability of returning to state $i$, Eq.~\eqref{eq:ReturnProb}, under neutrality becomes
\begin{align}\label{eq:Neutral03}
\begin{split}
	r_i=1-1/N. 
\end{split}
\end{align}
Hence, using Eq.~\eqref{eq:Sojourn03} the unconditional average sojourn times are
\begin{align}\label{eq:Neutral04}
\begin{split}
	t_{ij}=\begin{cases}
		N\frac{N-i}{N-j}\,\, \text{for}\,\, i>j\\
		N\frac{i}{j}\,\, \text{for}\,\, i\leq j.
	\end{cases}
\end{split}
\end{align}
The unconditional average fixation times follow from a summation over all $j$. 
With some basic algebra this summation leads to the unconditional average fixation time $t_i=N\, i \left(H_{N-1}-H_{i-1}\right)+N (N-i) \left(H_{N-1}-H_{N-i}\right)$, where we use the notation $H_k$ to indicate harmonic numbers, see definition in Table \ref{tab:MainSymbols}, or Eq.~\eq{HarmonicNumber} in the Appendix.   

The neutral conditional average fixation times can be found in a similar way, by multiplying with $\phi_j/\phi_i=j/i$, and summing over $j$, Eq.~\eq{Cond01}. 
The part of the sum with $i>j$ thus has terms of the form $\frac{j}{i}\frac{N-i}{N-j}$. In the other part of the sum with $i\leq j$, the ratios cancel, which gives a constant contribution. 
Terms in initial state $i$ factor out, such that we find
\begin{align}\label{eq:Neutral05}
\begin{split}
	t_i^N=\frac{N(N-i)}{i}\left( N(H_{N-1}-H_{N-i}) + 1\right). 
\end{split}
\end{align}
For a single mutant ($i=1)$ the conditional average fixation time under neutral conditions is thus $N^2-N$ time steps, or $N-1$ generations. 
Note that in fact neutral evolution means $T_i^\pm\leq 1/4$, so that the probability for the process to stay put, $1-T_i^+-T_i^-$, is greater than $1/2$, making the neutral process `lazy' in the terminology of \cite{montenegro:2006bo}). 

Neutral evolution serves as the reference baseline, and we ask how weak (but non-vanishing), intermediate, and strong selection changes the conditional fixation times with respect to this benchmark.
We focus our analysis on classical evolutionary games exhibiting counterintuitive behavior of fixation times as a function of selection intensity. For these cases we investigate weak and strong selection ($\beta\\ll1$, and $\beta\to\infty$, respectively). In these cases it is possible to derive closed-form analytical results for the conditional fixation times, and seemingly contradictory results may be found. For example the weak selection limit can indicate that small, but non-zero selection strength can lead to a slowdown of fixation, i.e., the conditional fixation times are higher than in the case of neutral evolution. On the other hand, a speedup effect can be seen for the same game in the limit of strong selection, indicating non-monotonic behavior of the conditional fixation time as a function of selection strength, a feature that we will investigate in more detail below.

 \subsection*{\sffamily Weak Selection}

The concept of weak selection is essential to many recent findings in evolutionary game theory \citep{nowak:2004pw,nowak:2006pw,traulsen:2006aa,ohtsuki:2006na,tarnita:2009df}. The weak selection limit here refers to a linearization of the effects of the game in terms of small values of the selection intensity $\beta$. In this limit the evolutionary game results in a small frequency dependent bias in addition to the undirected random drift of neutral evolution. Weak selection serves as a powerful first estimate of how selective bias and fluctuations in small populations influence each other. We can thus make use of a large body of existing results on fixation times. As one novelty we establish the weak selection approximation of the average sojourn times.  We do not present full mathematical details here, some of the main results can be found in Appendix \ref{sec:App02}.

Two relevant quantities for time scales in evolutionary games are given by the unconditional and conditional average fixation times of a single mutant,  $t_1$, and $t_1^N$. Remarkably, the fixation times under weak selection have a very simple dependence on the underling the evolutionary game \citep{altrock:2009nj}. One finds the weak selection approximations
\begin{align}
	t_1&\approx N\,H_{N-1}+\frac{\tilde E_2}{6}v\beta\label{eq:LinearWeakFT01},\\
	t_1^N&\approx N(N-1)-N(N-1)\frac{\hat E_1}{36}u\beta\label{eq:LinearWeakFT02},
\end{align}
where the game parameters enter in the distinct forms of Eqs.~\eq{PackageRule1}, and \eq{PackageRule2}. The coefficients $\tilde E_2$, and $\hat E_1$ in these expansions depend only on the population size $N$ and on the state $i$, but not on the game parameters. Detailed expressions are given in Appendix \ref{sec:App02}.

\subsection*{\sffamily Strong selection}

The specific choice we made for the mapping from payoffs to fitness, $f_A=\exp(\beta\pi_A)$, $f_B=\exp(\beta\pi_B)$, allows a strong selection limit; \cite{altrock:2009aa} have discussed strong selection ($\beta\to\infty$) for a closely related process with selection at birth and at death. 
We consider the limiting case of strong selection in order to compare it with the weak selection approximation.  
A natural question to ask is: Do weak selection approximation and strong selection limit agree qualitatively? As discussed below,
some games show a slowdown at small selection strength, the weak-selection limit predicts an increase of fixation time with increasing selection. On the other hand, in the strong selection limit of the same games the beneficial mutant can also fixate faster than in the neutral limit. Hence the time-to-fixation can be non-monotonic in the selection intensity.

The transition probabilities as defined in Eqs. (\ref{eq:Trans01a}, \ref{eq:Trans01b}), can be written as $T_i^\pm=i(N-i)/N^2\,g_i^{\pm}$, see also \citep{bladon:2010ws}. 
For the specific dynamics we study here, we have 
\begin{align}
	g_i^+&=\frac{1}{\frac{i}{N}(1-\e^{-\beta(\pi_A-\pi_B)})+\e^{-\beta(\pi_A-\pi_B)}},	\label{eq:Strong01A}	\\
	g_i^-&=\e^{-\beta(\pi_A-\pi_B)}\,g_i^+.	\label{eq:Strong01B}
\end{align}
As we are interested in the slowdown effect, we will now restrict the detailed analysis to cases in which strategy $A$ is always favored by selection, irrespective of the state $i$ of the population, i.e., to cases in which $\pi_A>\pi_B$, that is $\exp(-\beta(\pi_A-\pi_B))<1$.  
The other cases follow in a similar way, and we refer to the discussion in \citep{altrock:2009aa}. 
Taking the limit of infinite selection strength we obtain
\begin{align}
	\lim_{\beta\to\infty} &g_i^+=N/i,	\label{eq:Strong02A}\\
	\lim_{\beta\to\infty} &g_i^-=0,	\label{eq:Strong02B}
\end{align}
and we have 
\begin{align}
	\lim_{\beta\to\infty}&T_i^+=\frac{N-i}{N},	\label{eq:Strong03A}\\
	\lim_{\beta\to\infty}&T_i^-=0	\label{eq:Strong03B}
\end{align}
for the transition probabilities per time step. Hence, the $A$ mutant will always spread, but the process remains stochastic, as $T_i^+<1$ for all internal states. 
The strong-selection limit of the probability to ever reach state $j$, starting from $i$, is
\begin{align}\label{eq:Strong04}
\begin{split}
	\lim_{\beta\to\infty}\phi_{i j}=
	\begin{cases}
	1\,\,\text{if}\,\,i\leq j,\\
	0\,\,\text{else}.
	\end{cases}
\end{split}
\end{align}
Especially, fixation of $A$ occurs with probability one, $\phi_i\to1$, which implies $t_i^N\to t_i$. Nevertheless, because of non-vanishing waiting probabilities $T_i^0=1-T_i^+$, the fixation times still fluctuate. 
For the return probability, Eq.~\eq{ReturnProb}, we simply obtain $r_j=j/N$,  so that we have
\begin{align}\label{eq:Strong05}
\begin{split}
	\lim_{\beta\to\infty}t_{i j}=
	\begin{cases}
	\frac{N}{N-j}\,\,\text{if}\,\,i\leq j,\\
	0\,\,~~~~\text{else} 
	\end{cases}
\end{split}
\end{align}
 for the average sojourn times. The average fixation time is the sum over all sojourn times, which amounts to 
\begin{align}\label{eq:Strong06}
\begin{split}
	t_i=\sum_{j=i}^{N-1}\frac{N}{N-j}=N\,H_{N-i},
\end{split}
\end{align}
where $H_k$ are the harmonic numbers, see Table \ref{tab:MainSymbols}. 
Equation \eq{Strong06} is the asymptotic result for the unconditional and conditional average fixation time of advantageous $A$ mutants in the limit of strong selection. 
If $B$ is favored by selection instead, a similar analysis can be carried out, leading to $t_i=N(H_{N-1} - H_{N-i-1})$ in the limit of strong selection.

Coordination games have a more intricate dependency on the initial condition, but we note that the derivation is analogous to the cases of simple dominance above.
For coexistence games, the fixation times diverge with the selection intensity $\beta$, which can be shown by arguing that there exist an internal state $j_1$, and an adjacent internal  state $j_2=j_1+1$, such that $T_{i<j_1}^-\to0$, and $T_{i>j_2}^+\to0$ for all other internal states $i$ in the limit of strong selection.  
Hence, $r_{j_1},r_{j_2}\to1$, and the sojourn times $t_{ij_1}$ and $t_{ij_2}$ diverge, irrespective of the state $i$ from which the dynamics is started. Thus the fixation time diverges as well.

\subsection[Two Generic Examples]{Two Generic Examples: Prisoner's Dilemma and Coexistence}

Let us now reduce the analysis to games with fixed $a=0$, and $d=1$. The payoff matrix then reads
\begin{align}\label{eq:PmatrixEGS}
\bordermatrix{
& A & B \cr
A & 0 & b \cr
B & c & 1 \cr}
\end{align}
For $c<0$ and $b>1$, we have a Prisoner's Dilemma, in which the invading mutant $A$ corresponds to defection, and the wildtype $B$ corresponds to cooperation: 
The punishment for mutual defection is zero, the reward for mutual cooperation is one, both are kept fixed. 
The two free parameters of the game are then the temptation to defect against a cooperator, $b$, and the suckers payoff for cooperating with a defector, $c$. 
Defection dominates cooperation, for each frequency defectors have a higher payoff than cooperators. 
As defectors spread, the average fitness decreases monotonically. We will here be interested in the fixation of defectors in finite populations, starting from a situation in which there is one defector initially. Questions we will ask are then for example: Given that the defectors eventually take over, how long does this take, and how does it depend on the selection strength? 
As explained above, our benchmark when testing the effects of selection is the case of neutral evolution, $t_1^N=N^2-N$. Naively, one would expect that increasing the selection strength should enhance the evolutionary advantage of defectors, and hence reduce the time to fixation at the all-defect absorbing state. As we will see below, the behavior of the system is much more intricate, though. 

We start by analyzing the weak-selection limit. For the generic subset of games that fulfill $b+c>1$, the linear weak selection approximation predicts an increasing time to fixation of one initial defector as a function of selection strength. This can be seen from the analytical results presented in more detail in the Appendix, see in particular Eq.~\eq{WSTN}, where we note that $\hat E_2$ vanishes for a single initial defector. One then finds that the difference between the conditional fixation time at small selection intensity and that at neutrality is of the form $-C(N)u\beta$ (compare Eqs.~\eq{PackageRule1}, \eq{LinearWeakFT02}), where $c(N)$ is a constant which only depends on the population size, and where $C(N)>0$ for $N\geq3$, compare also to \citep{altrock:2009nj,taylor:2006jt}. 
As a consequence, the conditional fixation time increases under weak selection when $u=1-b-c<0$, i.e., when $b-c>1$, the conditional fixation time at low non-zero intensities of selection is then higher than in the neutral case.  In the limit of strong selection, on the other hand, we expect that the conditional fixation time approaches the value $N\,H_{N-1}$, Eq.~\eq{Strong06}, which is less than the neutral result of $N(N-1)$ (the harmonic numbers diverge logarithmically in $N$).  
This implies a non-monotonic behavior of $t_1^N$ as a function of $\beta$, with at least one maximum at intermediate values of the intensity of selection. 
\begin{figure}[t!]
\begin{center}
\includegraphics[width=0.75\linewidth]{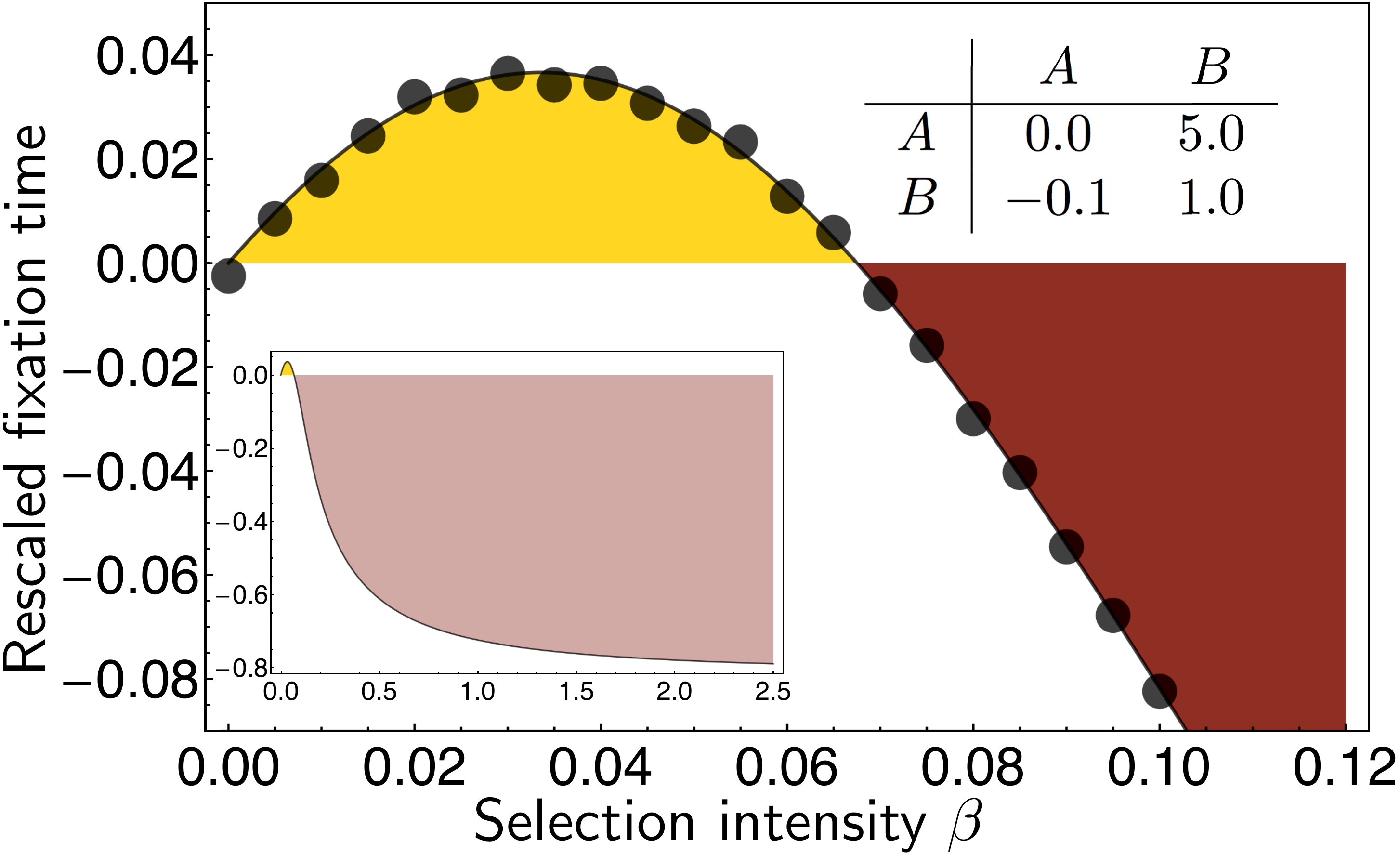}
\end{center}
\caption{\sffamily 
{\bf \sffamily Prisoner's Dilemma: Non-monotonic fixation time.} 
Average conditional fixation time of a single defecting mutant in a population of cooperators, relative to the case of neutral evolution (quantity shown is $t_1^N(\beta)/t_1^N(0)-1$) for $N=20$. 
The fixation time under neutral evolution is $t_1^{20}(0)=380$, see Eq.~\eq{Neutral05}. 
The main panel shows the result for small selection intensity, the inset depicts larger values of $\beta$. 
Solid lines are the semi-analytical solution, Eq.~\eq{Cond01}, grey dots are from Monte-Carlo simulations ($10^6$ independent realizations).  The fixation time approaches a value of  $N\,H_{N-1}$ for $\beta \to \infty$, Eq.~\eq{Strong06}. Accordingly, $t_1^N(\beta)/t_1^N(0)-1$ approaches $\approx -0.81$, as seen in the inset. }\label{fig:PDHump}
\end{figure}

In order to confirm this further, we plot the semi-analytical solution of the conditional fixation time from a single defector mutant, as obtained from explicitly carrying out the sum in Eq.~\eq{Cond01}, in Figure \ref{fig:PDHump}. For $c=-0.1$ and $b=5.0$, the conditional fixation time depends non-monotonically on the intensity of selection:
If the intensity of selection is roughly below $0.07$, fixation is expected to take longer than neutral. 
The analytical curve is corroborated by results from individual based Monte-Carlo simulations of the underlying birth-death process. It is interesting to note that while the behavior of $t_1^N(\beta)$ is nonlinear with an intermediate maximum, the actual slowdown, i.e., the initial increase of $t_1^N$ with the selection intensity $\beta$ occurs in the regime of weak selection. The maximum in fixation time is seen when selection intensity is of order $1/N$. 
The slowdown effect is only relatively small in our example (approximately $4\%$ relative to the neutral case). This is due to special features of the Moran process and to the numerical values we have chosen for the payoff matrix; the increase in fixation time can be much larger in other frequency dependent microscopic processes, as discussed in \citep{altrock:2010aa}. 
\begin{figure}[t!]
\begin{center}
\includegraphics[width=.95\linewidth]{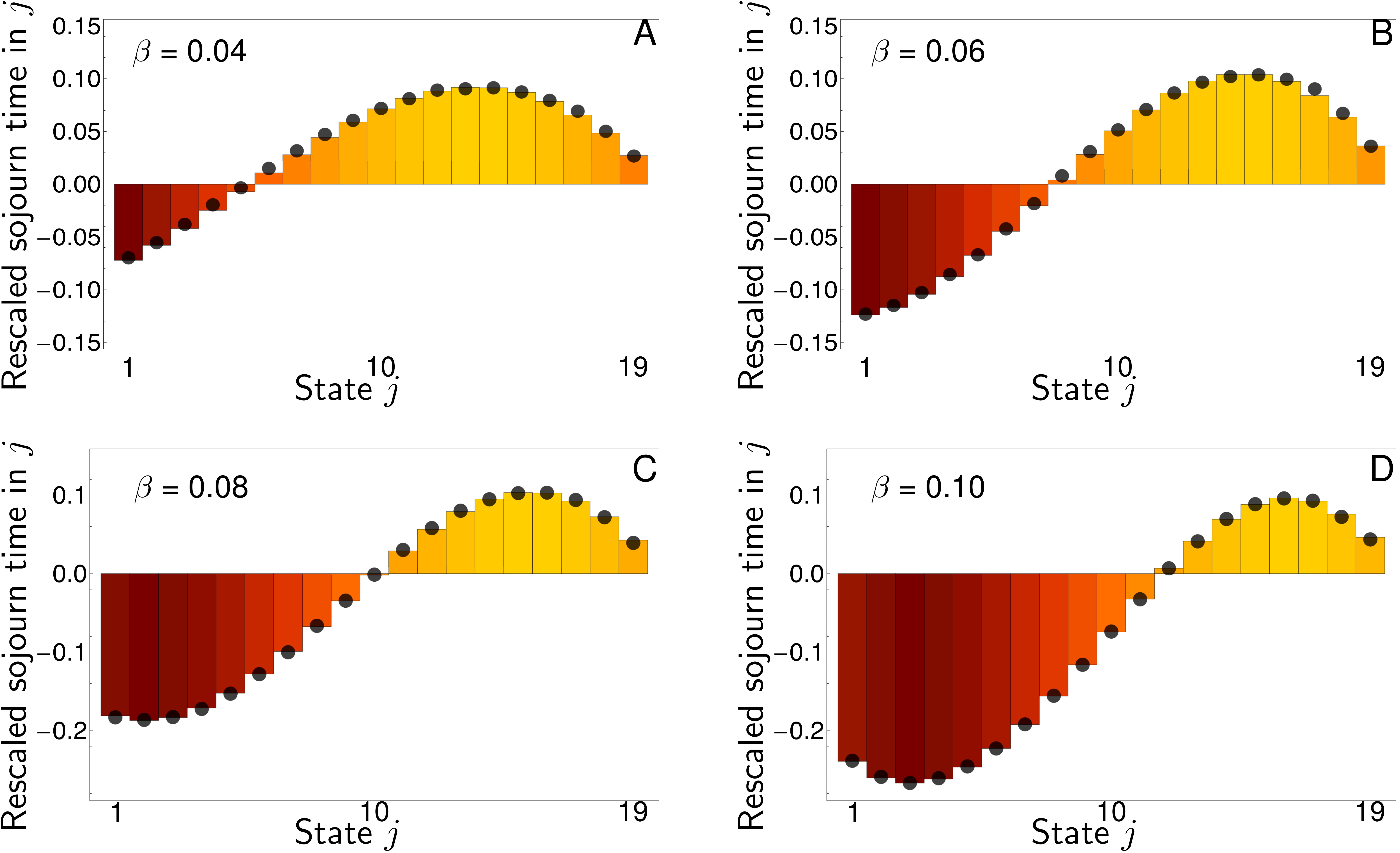}
\end{center}
\caption{\sffamily 
{\bf \sffamily Prisoner's Dilemma: Sojourn times.} 
The conditional average sojourn time in state $j=1,\dots,N-1$ of a single defecting mutant in a population of cooperators, 
rescaled relative to the neutral case (quantity shown is $t_{1\,j}^N(\beta)/t_{1\,j}^N(0)-1$) for four different values of selection intensity $\beta$ as indicated in each panel. 
Shaded bars refer to the exact analytical solution $\phi_j/\phi_i t_{1,j}$, dots are obtained from $10^6$ independent individual based Monte-Carlo simulations of the evolutionary Moran process. 
For small $\beta$, the total additional time spent in states close to $j=N$ exceeds the time gain in the states close to $j=0$ and slowdown occurs. 
Payoff values $a=0.0$, $b=5.0$, $c=-0.1$, $d=1.0$, population size $N=20$.  
}\label{fig:PDSoj}
\end{figure}

Slowdown in a social dilemma game, where cooperation is dominated by defection, is a phenomenon of small population size, it vanishes with increasing population size in the following sense: the larger the population, the smaller the range of selection intensities to observe slowdown. More specifically the range over which the slowdown occurs scales as $1/N$. This holds for the frequency-dependent Moran process, its variant studied here, imitation processes, as well as the Wright-Fisher process \citep{altrock:2009nj,wu:2010aa}. 
The non-monotonic behavior of average fixation times is influenced by the payoff structure of the game in a complex manner, and it is hard to predict for which parameter ranges slowdown is extreme. However, as shown in \citep{altrock:2010aa}, stochastic slowdown occurs in a variety of simple microscopic processes, and is facilitated in systems with only a small number of selective states and a large range of almost neutral states.

In order to understand the slowdown effect and the non-monotonicity of the time-to-fixation in more detail, we can ask where the process spends most of its time, i.e., we  consider the conditional sojourn times, $t_{1,j}^N$ for the intermediate states $j=2,3,\dots,N-1$. Results from our semi-analytical calculations and from simulations are shown in Figure \ref{fig:PDSoj}. As before we compare results from the selection scenario to those for neutral evolution. 
The neutral conditional sojourn time in each frequency $j$ does not depend on either of the two states, $i$ or $j$, but only on the population size $N$, $\lim_{\beta\to 0} t_{ij}^N=\lim_{\beta\to 0}[\phi_j/\phi_i\,t_{i\,j}]=N$.  
The reason is that the fixation probability in each state increases linearly in $j$, e.g., $\phi_j/\phi_i=j/i$. This compensates the asymmetry of the neutral unconditional sojourn times, which decrease inversely with increasing distance from initial condition, e.g., $t_{ij}=N\,i/j$.  

The independence on the intermediate states no longer holds for $\beta>0$, see Appendix \ref{sec:App02}. 
The non-monotonicity of the conditional fixation time of an advantageous defector results from a competition of two effects: 
We observe in Figure \ref{fig:PDSoj} that the system spends less time in states of low frequencies of defectors, $j$, than under neutral evolution, 
but that states of high frequencies are sojourned more often than in the neutral regime. The increase of the conditional sojourn times close to the absorbing boundary $i=N$ is a combined effect of high waiting probabilities (low transition rates $T_i^\pm$), and the fact that processes conditioned on fixation in $N$ are treated as having a reflecting boundary in $i=0$.  If selection is sufficiently weak, the reduction in sojourn times for states with low $j$ is relatively small, and only occurs for a few states. The majority of states experience an increased sojourn time relative to neutral. Thus, the initial gain in speed due to positive selection cannot compensate the later time loss. As the strength of selection increases, more and more states experience an effective reduction in sojourn times (see Figure \ref{fig:PDSoj}) and the relative time gain in each of these states also increases. An increased sojourn time relative to the neutral case is now only found near the $j=N$ state, so that the reduction at the beginning of the evolutionary trajectory now dominates, leading to a speedup. 
 
We point out that the seemingly counterintuitive slowdown and the non-monotonic behavior of $t_1^N$ as a function of $\beta$ rests in the fact that we condition on fixation in $N$. We consider only the ensemble of trajectories that eventually fixate in $N$, this ensemble naturally grows with increasing frequency dependent selection, but the average time it takes any of these trajectories to reach state $N$ shows complex behavior and the slowdown effect.

Next we discuss the behavior of $t_1^N(\beta)$ for the snowdrift game representing the class of coexistence games, $b>1$ and $c>0$. 
In these games, $A$ can invade $B$ and $B$ can invade $A$, a stable coexistence of the two strategies typically evolves. The snowdrift game is chosen frequently as a representative of this class \citep{doebeli:2004bo,doebeli:2005aa}:
Cooperators can be invaded by defectors as the temptation to defect is still larger than the reward of mutual cooperation. 
However, cooperating against a defector now yields a payoff greater than mutual defection, such that cooperation and defection can coexist in a population. 
The snowdrift game is a social dilemma, as selection does not favor the social optimum of mutual cooperation.

\begin{figure}[t!]
\begin{center}
\includegraphics[width=0.75\linewidth]{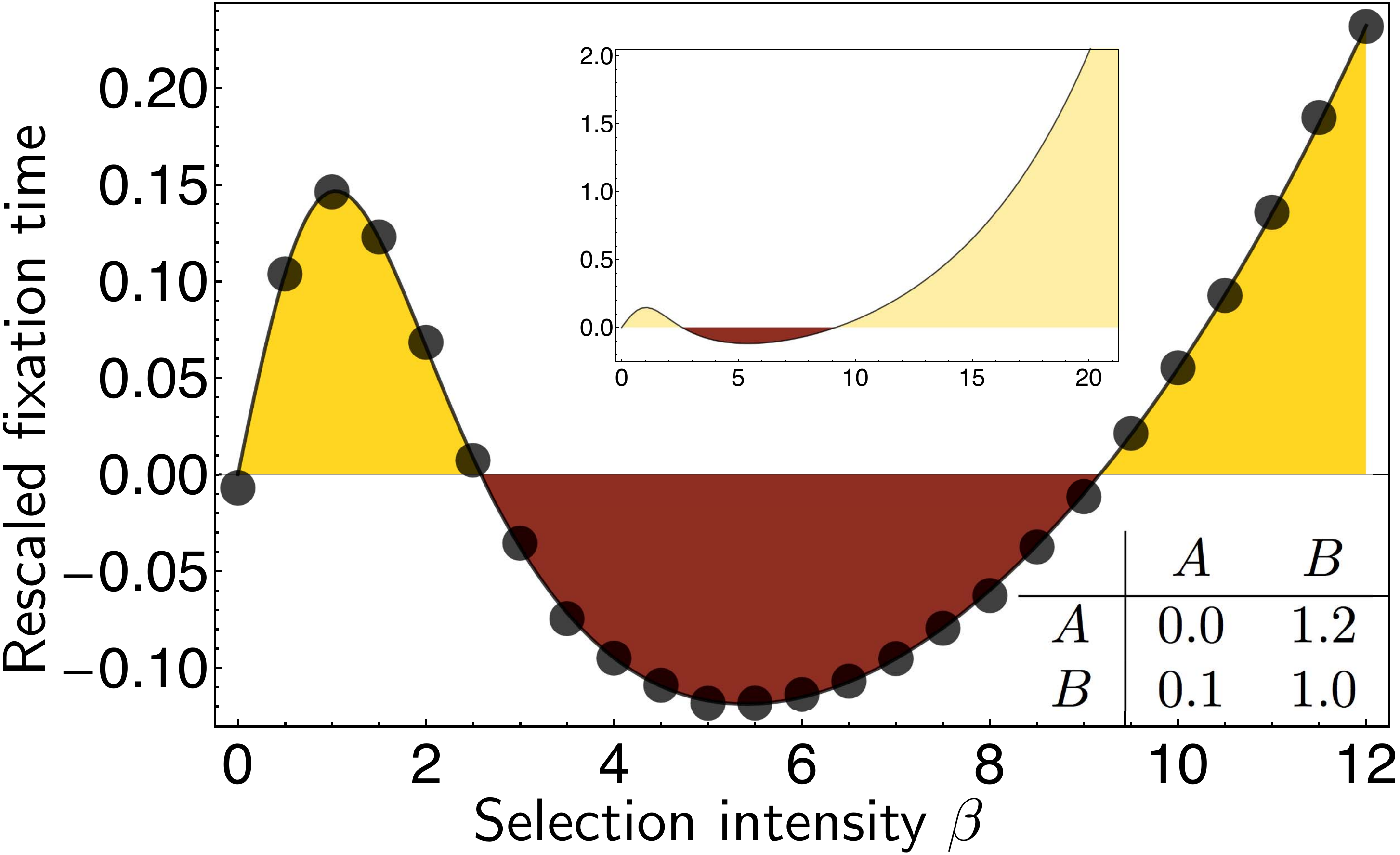}
\end{center}
\caption{\sffamily 
{\bf \sffamily Coexistence of $A$ and $B$: Non-monotonic fixation time.} 
Conditional average fixation time of a single mutant, measured relative to the neutral case (quantity plotted is $t_1^N(\beta)/t_1^N(0)-1$).  The fixation time under neutral evolution is $t_1^{30}(0)=870$, see Eq.~\eq{Neutral05}. The payoff values of the coexistence game are $a=0.0$, $b=1.2$, $c=0.1$, $d=1.0$, in a population of size $30$ individuals. The black lines are the exact analytical solution, Eq.~\eq{Cond01}, grey dots are from individual based Monte-Carlo simulations ($5\cdot 10^5$ independent realizations).    
{\bf \sffamily Inset:} For very strong selection, the fixation time diverges. 
}\label{fig:CoexHump}
\end{figure}

Again, we denote wildtypes $B$ as cooperators and the mutants $A$ as defectors. In coexistence games, the frequency dependent fitness difference $f_A-f_B$ is positive at low numbers of cooperators. A population consisting mostly of cooperators can hence be invaded by defectors. Once the number of defectors has reached a certain threshold, the fitness advantage is reversed, $f_A-f_B$ becomes negative. In the corresponding deterministic replicator equation, this leads to a stable fixed point at which both types coexist. The stochastic dynamics of finite populations of individuals interacting in coexistence games will still fixate into the absorbing all-$A$ or all-$B$ states eventually. But due to the deterministic flux towards the coexistence fixed point, the time-to-fixation diverges rapidly with selection pressure \citep{traulsen:2007cc}. Our numerical analysis, however, reveals that there are generic subsets of coexistence games in which the conditional fixation time can decrease for intermediate values of selection, as illustrated in the example of Figure \ref{fig:CoexHump}. 

\begin{figure}[t!]
\begin{center}
\includegraphics[width=.95\linewidth]{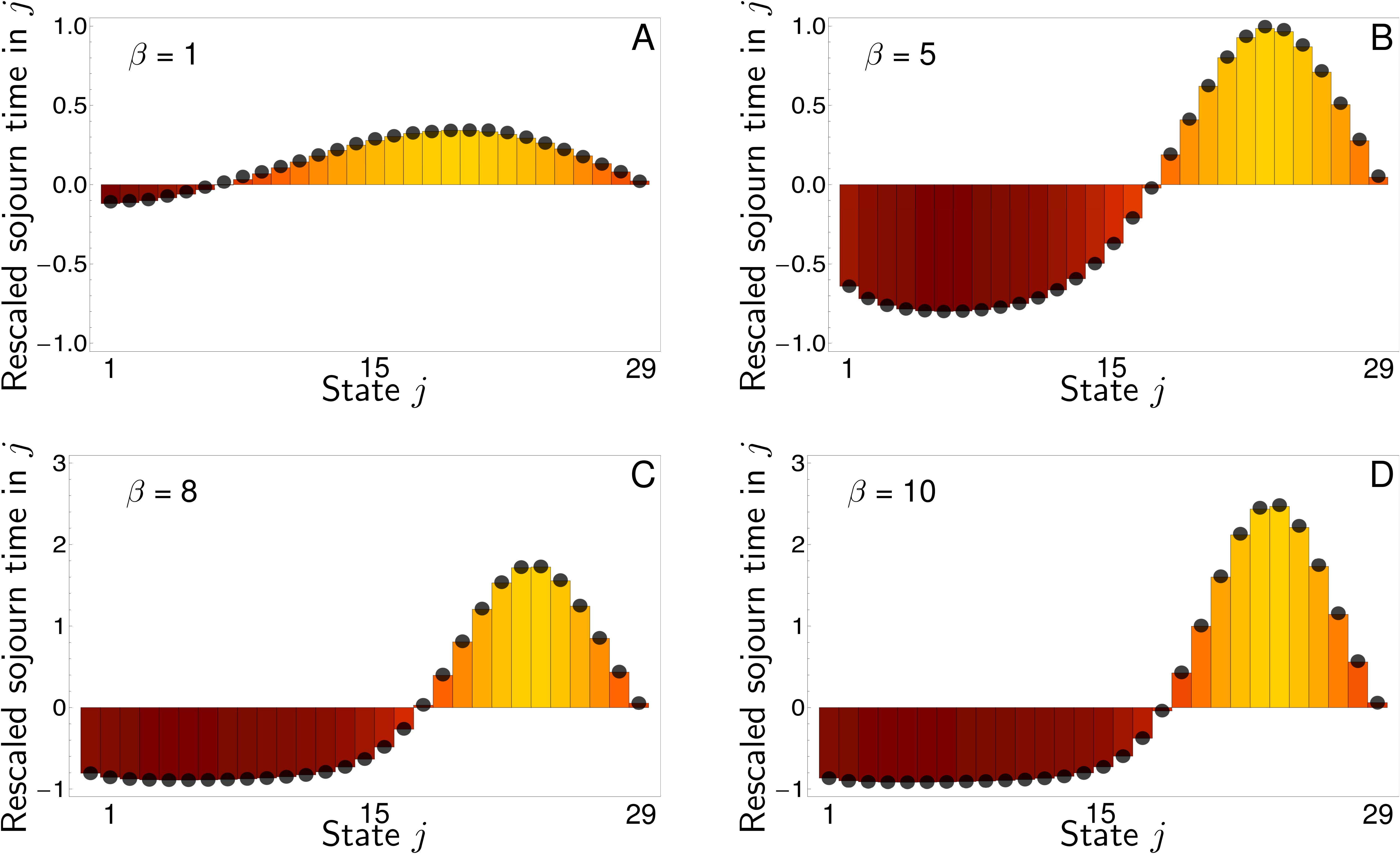}
\end{center}
\caption{\sffamily 
{\bf \sffamily Coexistence of $A$ and $B$: Sojourn times.} 
Conditional average sojourn time in state $j=1,\dots,N-1$ of a single defecting mutant in a population of cooperators,  relative to the neutral case (quantity shown is $t_{1\,j}^N(\beta)/t_{1\,j}^N(0)-1$), using four different values of the selection intensity $\beta$ as indicated in the four panels.  
Shaded bars refer to the exact analytical solution $\phi_j/\phi_i t_{1,j}$, dots are obtained from $10^5$ independent individual based Monte-Carlo simulations of the evolutionary Moran process.
Payoff values are $a=0.0$, $b=1.2$, $c=0.1$, $d=1.0$, population size $N=30$.  
}\label{fig:CoexSoj}
\end{figure}

The dependence of $t_1^N(\beta)$ on $\beta$ is surprisingly intricate. The conditional fixation time $t_1^N$ shows two extrema as a function of the selection strength. 
In order to present a more complete picture we complement the calculation of fixation times by results for conditional sojourn times in Figure \ref{fig:CoexSoj} for several values of the selection pressure. The typical picture is that, at non-zero selection strength, states of low frequency of $A$ are visited substantially less often than under neutrality, but the fixation process spends long periods at high-frequency states.  At low (but non-zero) selection strength the number of states near $j=1$ that experience a reduced sojourn time is small (see Figure \ref{fig:CoexSoj}). The majority of states is subject to an increase in sojourn time relative to the neutral case, leading to an overall increase the fixation time. At intermediate strength of selection a sizeable fraction of states experiences a substantial reduction of sojourn times (see e.g., the case $\beta=5$ in Figure \ref{fig:CoexSoj}), this effect dominates over the effective slowdown in states near $j=N$. The net effect of these two competing forces is then an overall speedup. Unlike in the case of the Prisoner's Dilemma however, in which virtually all states experience a reduction in sojourn time at strong selection (see Figure \ref{fig:PDSoj}), the set of states in which sojourn times are reduced is limited in the coexistence game studied in Figure \ref{fig:CoexSoj}. Even for very large values of $\beta\approx 20$ we find that sojourn times are increased compared to neutral for states $j$ approximately above $j=20$ in a population of $N=30$. At the same time the sojourn time in these states near $j=N$ increases considerably as $\beta$ is increased (see lower right panel of Figure \ref{fig:CoexSoj}), leading to an overall slowdown effect, and ultimately to a conditional fixation time which diverges with $\beta$.

\subsection*{\sffamily Analysis of Generic $2\times 2$ Games}

\begin{figure}[b!]
\begin{center}
\includegraphics[width=.65\linewidth]{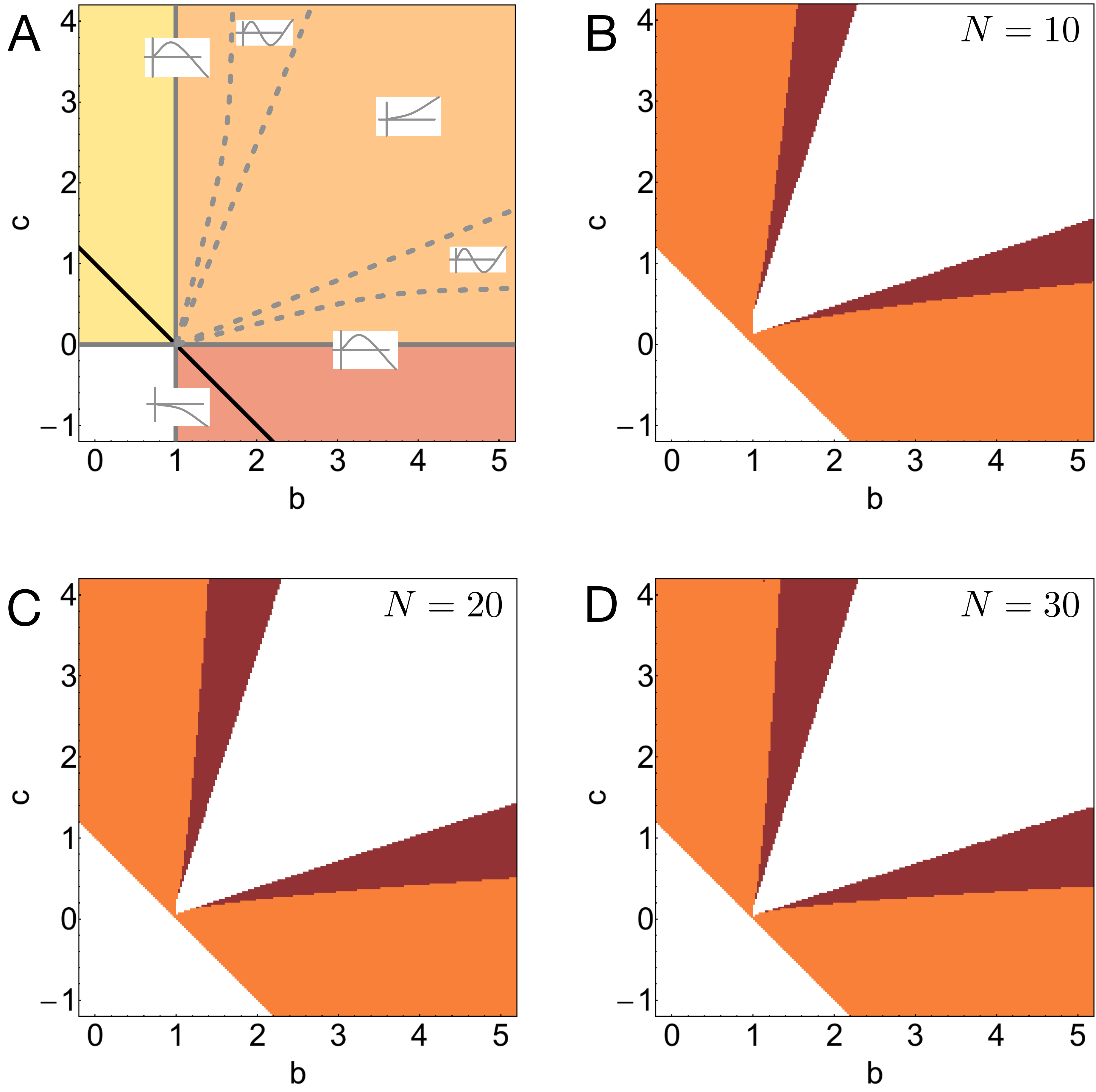}
\end{center}
\caption{\sffamily 
{\bf \sffamily Phases of possible non-monotonic behavior of the fixation time $t_1^N$, as a function of $\beta$ in the game plane parameterized by $b$ and $c$.} 
Panel A shows a schematic drawing of the parameter space, the small inset figures depict the qualitative shape of the conditional fixation time $t_1^N$ as a function of the selection strength $\beta$. The parameters plane can be split into four quadrants separated by the vertical line of $b=1$, and the horizontal line of $c=0$. The lower left quadrant defines coordination games. The lower right quadrant defines a Prisoner's Dilemma game with dominance of $A$ over $B$. In the upper right quadrant, strategies $A$, and $B$ define a coexistence game, in the upper left one, $B$ dominates (sometimes called `harmony game'). The black line $c=1-b$ (along which all games have the equal-gains-from-switching property) separates the two regimes in which the conditional fixation time $t_1^N$ increases (above), or decreases (below) with increasing linear weak selection. In the coexistence quadrant we schematically draw dashed lines that separate different regimes of non-monotonic behavior of the conditional fixation time $t_1^N$ with selection intensity $\beta$. The three following panels B, C, and D scan the $b-c$ plane for the number of non-trivial solutions of $t_1^N(\beta)=t_1^N(0)$; Bright shading means one solution (hence one extreme value), dark shading means two solutions (hence two extreme values) for finite $\beta$. In the white regions $t_1^N(\beta)$ has no extrema as a function of $\beta$. In the three examples shown in panels B, C and D, the parameter area in which a coexistence game has two extrema in $t_1^N(\beta)$ increases as $N$ is increased.  
}\label{fig:bc_Plane}
\end{figure}
Here we describe the behavior of more general $2\times 2$ games, parametrized by $b$, and $c$, with fixed $a=0$, and $d=1$. This results in a parameter plane, spanned by $b$ and $c$. While this is a reduction of the space of all possible games, all generic types of $2\times 2$ evolutionary games are captured by this parameterization \citep{weibull:1995hp}. In particular one has a coordination game for  $b<1$, $c<0$, a Prisoner's Dilemma type game for $b>1, c<0$ and a harmony game for $b<1,c>0$. 
The region $b>1, c>0$ represents coexistence games. 
We now ask whether the conditional fixation time $t_1^N$ as a function of $\beta$ has nontrivial solutions of $t_1^N(\beta)=t_1^N(0)$. 
If there are such solutions, then assuming that $t_1^N$ is smooth, there has to be at least one extremum. For systems with slowdown, the single extremum is a maximum. 
In cases where one finds speed-up there can be two extrema, a maximum and a minimum. 

We are interested in the type of complexity the resulting curves show depending on $b$, and $c$, i.e., how many extrema one should expect as a function of selection strength. In what follows we will discuss the generic classes of $2\times2$ games separately. 

\textit{Coordination games ($b<1$, and $c<0$):} 
In this case, our numerical analysis reveals, that no maxima or minima in the fixation time are possible. 
However, there is a finite strong selection limit, lower than that of neutral evolution. Hence the fixation time can be assumed to decrease monotonically with increasing selection. 

\textit{Prisoner's Dilemma ($b>1,c<0$), and Harmony Game ($b<1,c>0$)}:
Games with $b>1$, and $c<0$ are called social dilemmas because the social optimum is not the dominant strategy. Complementary, the dilemma is resolved if $b<1$, and $c>0$, where mutual cooperation is dominant. In both cases, we can observe slowdown for sufficiently weak selection, followed by a speedup for strong selection. This is seen if $b+c>1$, otherwise the conditional time-to-fixation simply decreases in $\beta$. In both quadrants of the parameter plane, the line that formally separates slowdown games from non-slowdown games is given by the set of games with equal gains from switching, $c=1-b$.

\textit{Coexistence games ($b>1$, and $c>0$):}
Here, fixation times have always been believed to diverge quickly with increasing selection. 
Our previous discussion reveals that for intermediate levels of selection, there is the possibility of a speedup, although initially, the fixation time increases as predicted by the linear weak selection approximation. 
Figure \ref{fig:bc_Plane} shows in which parts of the coexistence regime there can be a strongly non-monotonic dependence of the conditional fixation times on the intensity of selection, as shown in the example of Figure \ref{fig:CoexHump}. 

In accordance with the scaling analysis in \citep{altrock:2010aa}, non-monotonic behavior of $t_1^N$ that leads to the pattern in Figure \ref{fig:bc_Plane} vanishes with increasing population size.  For large populations, the range of the selection strength in which slowdown is found scales as $1/N$. 

Note also that in Figure \ref{fig:bc_Plane} (B-D) the regime of slowdown, to be observed in games where one strategy dominates the other, reaches into the game parameter area of coexistence games. 
This is due to the observation that in finite populations, not every game in which the deterministic dynamics predicts an mixed equilibrium is in fact a coexistence game. 
There is a finite size correction to the equilibrium value. 
If the equilibrium is below $1/N$, or above $1-1/N$, a formal coexistence game rather describes dominance in a finite population.

\section{Conclusions}\label{sec:conc}

The study of cooperation in an evolutionary context mainly focuses on mechanism that allow the emergence and maintenance of cooperation \citep[see e.g.,][]{axelrod:Science:1981,maynard-smith:book:1982,nowak:2006pw,santos:2006pn,sigmund:2010aa,hilbe:SciRep:2012,van-veelen:PNAS:2012}. 
In well-mixed populations, cooperation cannot be maintained when defection emerges unless mechanisms promoting cooperation are present, such as repetition, punishment, or rewarding. 
Nonetheless, in small groups of cooperators that are bound to be invaded by defectors stochastic effects can have a beneficial impact \citep{cremer:SciRep:2012}, leading to a possible delay of the extinction of cooperation. 
\cite{nowak:JTB:2012} summarizes important steps in the understanding of cooperation from an evolutionary perspective, and also points out future challenges in the experimental study of human cooperation. 
Our analysis here can be seen as a next step towards a more intuitive understanding of large fluctuations, the time a dominated strategy can be expected to be maintained, and the importance of the particular game that is being played to overall fitness.

In summary we have studied the time-to-fixation in finite populations of individuals interacting in $2\times 2$ games, subject to a birth-death process. In particular we have focused on the detailed mechanics of a stochastic slowdown effect, previously reported for simple evolutionary processes in \citep{altrock:2010aa}. The term slowdown refers to cases of non-zero selection strength in which fixation of a single advantageous mutant in a population takes longer on average than in the neutral case. Our analysis proceeds partly based on numerical simulations and partly by exact mathematics, the latter resulting in closed-form expressions for fixation times which are then evaluated numerically. We derive the weak and strong selection limits of these relations. Considerations of a simple toy model complement our mathematical analysis of the more intricate and complex games. For simple-birth death processes we can identify a small set of model parameters contributing to the time-to-fixation, and hence the competing effects leading to slowdown or speedup can be disentangled.

Depending on the specific choice of the payoff matrix we find intricate functional dependence of the fixation time on the intensity of selection, with non-monotonicities and multiple intermediate extrema. Generic instances of the Prisoner's Dilemma may exhibit an initial slowdown of fixation at low, but non-zero selection strength, followed by a speedup into a faster-than-neutral regime at strong selection. In coexistence games we have identified an initial slowdown at low selection, followed by a speedup at intermediate intensity of selection, followed again by a slowdown. In order to systematize these different types of behavior we have classified the parameter plane of $2\times 2$ games, characterized by two generic game parameters, according to the complexity their respective conditional fixation times show as a function of selection strength. 

As a second main contribution we have linked fixation times and the slowdown effect to average sojourn times, i.e., to the time the system spends in different states before fixation into the all-mutant absorbing state occurs. We find that the slowdown (relative to the case of neutral evolution) is caused by an increase of the time spent near the all-mutant state. The time spent near the all-wildtype state is reduced increasingly as selection strength sets in. Both effects compete, and can result in a non-monotonic behavior of the conditional fixation time, such as the one shown in Figure \ref{fig:PDHump}.   

While we focus on $2\times 2$ games and restrict the discussion to a variant of the Moran process, we expect the slowdown effect and possible non-monotonic behavior of conditional fixation times to play a role in other systems as well. The connection between the evolutionary game with payoff matrix \eq{Pmatrix}, and the evolutionary dynamics of two alleles, $A$ and $B$, at a single locus in a population of diploid organisms with random pairing of gametes is, for example, discussed by \cite{traulsen:JTB:2012}. Given these parallels it will not be surprising if a stochastic slowdown is to occur in model systems in population genetics with directional selection. 

While we appreciate that slowdown effects may not have been found in real-world experimental systems so far, we would like to argue that -- up to recently -- there has not been much reason to look for them systematically. We hope that our theoretical findings may stimulate a discussion of how long a beneficial mutation needs to reach fixation in a population, especially in systems for which there exist estimates of selection strength and fitness function. With such systems in mind, we hope that it may then be possible to test the applicability of our findings in real-world systems.

We would like to conclude by briefly addressing the symmetry in conditional fixation times, discovered by \cite{antal:2006aa} and \cite{taylor:2006jt}: One has $t_1^N=t_{N-1}^0$ for all $2\times2$ games in finite populations with two strategies $A$ and $B$, for all microscopic birth-death dynamics and any selection strength (excluding pathological cases in which some of the transition rates vanish). The conditional average fixation time of a single $A$ individual is always the same as that of a single $B$ individual.
Given this rather surprising symmetry it is natural so ask whether conditional sojourn times have a similar property, i.e., whether $\phi_j/\phi_1 t_{1 j}=(1-\phi_j)/(1-\phi_1) t_{N\!-\!1 j}$? 
Here, a numerical comparison suggests that this is indeed the case, see also \citep{taylor:2006jt}.

\vspace{2em}
{\bf \sffamily Acknowledgements}\\
TG is grateful for funding by the Research Councils UK (RCUK reference EP/E500048/1), and by EPSRC (references EP/I005765/1 and EP/I019200/1). TG acknowledges hospitality by the Max-Planck-Institute for Evolutionary Biology, Pl\"on, Germany.   
PMA and AT acknowledge financial support from the Deutsche Forschungsgemeinschaft and the Max-Planck-Society. 
We gratefully acknowledge discussions with Bin Wu. 
\vspace{2em}

{\bf \sffamily Appendix}
\begin{appendix}

\section{Stationary Probabilities}
\label{sec:App01}

For the probability to ever visit state $j$ before absorption, starting from $i$, the recursion 
$\phi_{i j}=(1-T_{i}^+-T_{i}^-)\phi_{i j}+T_{ i}^+\,\phi_{i+1 j}+T_{i}^-\,\phi_{i-1 j}$
holds. 
For $i=j$, we simply find $\phi_{ i i}=1$. 
For $i>j$, the solution of the recursion with absorbing boundaries $0$, $N$ is \citep{ewens:2004qe} \begin{align}\label{eq:AppProb01}
	\phi_{i j}=\frac{
	\sum _{k=i}^{N-1} \prod _{m=j+1}^k \frac{T_m^-}{T_m^+}
	}{
	\sum _{k=j}^{N-1} \prod_{m=j+1}^k \frac{T_m^-}{T_m^+}
	}.
\end{align}
For $i<j$, solving the recursion leads to
\begin{align}\label{eq:AppProb02}
	\phi_{i j}=\frac{
	\sum _{k=0}^{i-1} \prod _{m=1}^k \frac{T_m^-}{T_m^+}
	}{
	\sum _{k=0}^{j-1} \prod_{m=1}^k \frac{T_m^-}{T_m^+}
	},
\end{align}
which yields the fixation probability $\phi_i=\phi_{ i N}$ as special case. Note that without mutation rates absorption is the only possible outcome, $\phi_{i 0}=1-\phi_i$.

\section{Weak selection}\label{sec:App02} 

The transition rates in state $i$ up to first order in $\beta$ read
\begin{align}
	T_i^+\approx&\,\frac{i(N-i)}{N^2}+\frac{i(N-i)^2}{N^3}(u\,i+v)\beta
		\label{eq:WSTa},\\
	T_i^-\approx&\,\frac{i(N-i)}{N^2}-\frac{i^2(N-i)}{N^3}(u\,i+v)\beta
		\label{eq:WSTb}.
\end{align}
For $i>j$, we obtain 
\begin{align}\label{eq:WSAlphaLo1}
\begin{split}
	\phi_{ij}\approx\,\frac{N-i}{N-j}\left(1-(i-j)\frac{(i+j+N)\,u+3\,v}{6}\beta\right). 
\end{split}
\end{align}
For $i<j$, we find
\begin{align}\label{eq:WSAlphaHi1}
\begin{split}
	\phi_{ij}\approx\,\frac{i}{j}\left( 1+(j-i)\frac{(i+j) u+3 v}{6}\beta\right),
\end{split}
\end{align}
which recovers the $1/3$-rule for $i=1$ and $j=N$ \citep{nowak:2004pw,lessard:2007aa}. 
The return probability in state $j$, before absorption, is given by $r_j=1+T_j^+(\phi_{j+1 j}-1)+T_j^-(\phi_{j-1 j}-1)$. The sojourn time in $j$ on the other hand is related to the escape probability $r_j$ from $j$, see Eq.~\eqref{eq:Sojourn03}.  Up to first order in the selection intensity we find
\begin{align}\label{eq:WSreturn01}
\begin{split}
	\frac{1}{1-r_j}\approx\, N+\frac{(N + j^2 (N+6)- N\,j(N+3) ) u -3 (N- 2j) v}{6}\beta. 
\end{split}
\end{align}
With this, and using Eq. \eqref{eq:Sojourn03} as well as Eqs. \eqref{eq:WSAlphaLo1} and \eqref{eq:WSAlphaHi1} we can compute the linear weak selection approximation for the average sojourn time in $j$. 
For initial conditions $i>j$ this reads
\begin{align}\label{eq:WSSojLo}
\begin{split}
	t_{i j}\approx\,\frac{N-i}{N-j}
		\left(N-\left(\hat D_1\,u+\hat D_2\,v\right)\beta\right),
\end{split}
\end{align}
where the quantities 
$\hat D_1=(i N^2+(i^2-1 + j(3 - 2 j))N-6 j^2)/6$ and 
$\hat D_2 = ((i-j+1)N-2j)/2$ are independent of the selection intensity and the payoffs.
 
For initial conditions $i<j$ we obtain
\begin{align}\label{eq:WSSoj}
	t_{i j}\approx\,\frac{i}{j}
		\left(N-\left(\check D_1\,u+\hat D_2\,v\right)\beta\right),
\end{align}
where $\check D_1= \hat D_1+(N^2(j-i))/6$. This enables us to calculate the average life time of the Markov chain between $0$ and $N$ under weak selection
\begin{align}\label{eq:WSUT}
\begin{split}
	t_i=&\sum\limits_{j=1}^{N-1}t_{i j}\\
	\approx&\, N\, i \left(H_{N-1}-H_{i-1}\right)\,+N (N-i) \left(H_{N-1}-H_{N-i}\right)\,+\left( \tilde E_1\,u + \tilde E_2\,v \right)\frac{\beta}{6}.
\end{split}	
\end{align}
Here, we have written
\begin{align}\label{eq:HarmonicNumber}
	H_k=\sum_{l=1}^{k}\frac{1}{l}
\end{align}
for the harmonic numbers.
The coefficients in Eq. \eqref{eq:WSUT} are given by
\begin{align}
	\tilde E_1=&\,\nonumber N(N-i)(N-3)(i-1)
	-iN(i^2-1)(H_{N-1}-H_{i-1})\nonumber\\
	&+(N-i)\Bigg( (i-1)(i (N+3) + N ( 2 N+3))\nonumber\\ 
	&- N(N-i+1) (2N + i+ 1)(H_{N-1}-H_{N-i}) \Bigg),\nonumber\\
	\tilde E_2=&\,i (N-i) 3(N+2) - 3iN(i+1)(H_{N-1}-H_{i-1})\nonumber\\
	&- (N-i) ((i-1) 3( N+2) - 18 N (N-i+1) (H_{N-1}-H_{N-i})\nonumber.
\end{align}

For $i=1$, the term $\tilde E_1$ vanishes for all $N$ such that the term proportional to $u$ drop out in Eq. \eqref{eq:WSUT}. 
Thus the average life time of a single mutant is given by $t_1\approx N H_{N-1}+ (N^2+N-2(1+N\,H_{N-1}))\,v\,\beta/12$, 
compare with the result by \cite{altrock:2009nj}. 

In a similar fashion, we can derive a weak-selection approximation for the conditional average fixation time
\begin{align}\label{eq:WSTN}
\begin{split}
	t_i^N\approx&\,\frac{N(N-i)}{i}\left( 1+N(H_{N-1}-H_{N-i}) \right)\\
	&+\frac{N(N-i)}{i}\left( \frac{\hat E_1}{36}u + \frac{\hat E_2}{2}v \right)\beta,
\end{split}
\end{align}
where
\begin{align}
	\hat E_1 =&\, (6 - 7 i) N^2+2 (9 + (i-9) i) N - 6 (i^2-1)\nonumber\\
	 &+6 N (1 + N (N - i + 3))(H_{N-1}-H_{N-i}),\nonumber,\\
	 \hat E_2 =&\, N(H_{N-1}-H_{N-i})-i+1  \nonumber.
\end{align}
The term $\hat E_2$ vanishes for all $N$ if $i=1$, and we arrive at the approximation $t_1^N\approx N(N-1)(1-(N^2-2N))\,u\,\beta/36$ 
\citep{taylor:2006jt,altrock:2009nj,altrock:2010aa}. 
The conditional extinction time of an initial group of $i$ mutants under weak selection is approximately
\begin{align}\label{eq:WST0}
\begin{split}
	t_i^0\approx&\,\frac{N \left(N\,i( H_{N-1}-H_{i-1})-N+i\right)}{N-i}\\
	&+\frac{N}{N-i}\left( \frac{\check E_1}{36} u + \frac{\check E_2}{2} v \right)\beta.
\end{split}
\end{align}
Here
\begin{align}
	\check E_1 =&\,6Ni(Ni+1)(H_{N-1}-H_{i-1})\nonumber\\
	&-(N-i)( 5 i N^2+ 2 i (3 + i)N - 6 (i^2 -1))\nonumber,\\
	\check E_2 =&\,(N-i)(i+1)-Ni(H_{N-1}-H_{i-1}).
\end{align}
We see that for $i=N-1$, $\check E_2$ vanishes, and $\check E_1(i=N-1)=\hat E_1(i=1)$, hence $t_{N-1}^0=t_1^N$, 
which holds for any intensity of selection \citep{taylor:2006jt,antal:2006aa}.

\end{appendix}

\bibliographystyle{elsarticle-harv}

\end{document}